\documentclass[preprint,tightenlines,floats,aps,amsmath,amssymb,nofootinbib,prd,showpacs,superscriptaddress]{revtex4}
\usepackage{graphicx} 
\usepackage{subfigure}
\usepackage{dcolumn} 
\usepackage{bm}
\usepackage{hyperref}

\begin{document}

\title{Constraining the nuclear symmetry energy and properties of neutron star from GW170817 by Bayesian analysis}

\author{Yuxi Li}
\affiliation{School of Physics and Optoelectronics, South China University of
	Technology, Guangzhou 510641, P.R. China}
\author{Houyuan Chen}
\affiliation{School of Physics and Optoelectronics, South China University of
	Technology, Guangzhou 510641, P.R. China}
\affiliation{School of Physics and Astronomy,
	Sun Sat-Sen University, Zhuhai 519082, P.R. China}
\author{ Dehua Wen\footnote{Corresponding author. wendehua@scut.edu.cn}}
\affiliation{School of Physics and Optoelectronics, South China University of
	Technology, Guangzhou 510641, P.R. China}
\author{Jing Zhang}
\affiliation{School of Physics and Optoelectronics, South China University of
	Technology, Guangzhou 510641, P.R. China}

\date{\today}

\begin{abstract}

Based on the distribution of tidal deformabilities and component masses of binary neutron star merger GW170817, the parametric equation of states (EOS) are employed to probe the nuclear symmetry energy and the properties of neutron star. To obtain a proper distribution of the parameters of the EOS that is consistent with the observation, Bayesian analysis is used and the constraints of causality and maximum mass are considered. From this analysis, it is found that the symmetry energy at twice the saturation density of nuclear matter can be constrained within $E_{sym}(2{\rho_{0}})$ = $34.5^{+20.5}_{-2.3}$ MeV at  90\% credible level. Moreover, the constraints on the radii and dimensionless tidal deformabilities of  canonical neutron stars are also demonstrated through this analysis, and the corresponding constraints are 10.80 km $< R_{1.4} <$ 13.20 km and $133 < \Lambda_{1.4} < 686$ at 90\% credible level, with the most probable value of $\bar{R}_{1.4}$ = 12.60 km and $\bar{\Lambda}_{1.4}$ = 500, respectively. With respect to the prior, our result (posterior result) prefers a softer EOS, corresponding to a lower expected value of symmetry energy, a smaller radius and a smaller tidal deformability.


\end{abstract}

\maketitle

\section{Introduction}

On August 17, 2017, the Advanced LIGO and Advanced Virgo first observed the merger of two neutron stars GW170817 \cite{Abbott 1}. Through continual research based on the data of this observation, the understanding of properties of neutron stars (such as the radius, the tidal deformability, etc.) and the state of dense nuclear matter is improved continually  \cite{Abbott a,Margalit b,Bauswein c,Abbott d,Baym 2,Abbott 6}.
Considering the case that the detection of gravitational radiation from the coalescence of a neutron star binary system is occasional \cite{Abbott 1,Abbott2020}, Bayesian inference becomes a popular method to analyze the observational data. In fact,  the Bayesian analysis is frequently used to investigate the properties and the state of the compact-star-matter in recent years \cite{Lim,De,Carson,Kastaun2019,Carreau,Lim 2,Lim 3,Carreau 2,Hernandez,Fasano,Riley}.

 Based on Bayesian analysis with equation of states described by chiral effective field theory,  Lim $et~ al.$  constrained the dimensionless tidal deformability of a 1.4 $M_{\odot}$ neutron star in a range of 136 $\leq$ $\Lambda_{1.4}$ $\leq$ 519 at 95\% credible level. Moreover, they  found an empirical relation between the tidal deformability of a canonical neutron star  and the pressure at twice nuclear saturation density, which provides a useful clue to investigate the state of the dense nuclear matter \cite{Lim}. By performing Bayesian analysis with the distance and source location derived by electromagnetic observations of GW170817 event, De $et~ al.$  constrained the $\tilde{\Lambda}$ ($\tilde{\Lambda}$ is defined as equation 3 in \cite{De}) at 90\% credible level as follows:  84 $\leq$ $\tilde{\Lambda}$ $\leq$ 642 for uniform component mass prior, 94 $\leq$ $\tilde{\Lambda}$ $\leq$ 698 for the distribution of component mass prior deduced from radio observations of Galactic binary neutron stars and 89 $\leq$ $\tilde{\Lambda}$ $\leq$ 681 for a component mass prior derived by radio pulsars \cite{De}.

As we know, the density of the neutron star matter covers a large range of magnitude, from a density far lower than the saturation density at the outer crust to a density close to 10 times the saturation density at the stellar center. At present, there is relatively small discrepancy in the EOS at density near or lower than the saturation density. But for the matter in the core with supra-saturation density, the EOS  is far from certain. In nuclear theory, there are too many EOS predictions based on various nuclear theories by using different interactions, and the predicted EOS often diverge at the supra-saturation density.
In fact, the uncertainty of the symmetry energy at supra-saturation density   is the main factor  leading to the divergency of the EOS \cite{Li}.
With the aid of the astronomical observations, people find a practice way to narrow the divergency. For example, by using the representative stellar radius data of canonical neutron star, Xie and Li \cite{Xie2019} inferred the high-density nuclear symmetry energy through  Bayesian inference by employing an isospin-dependent parametric EOS model for neutron star matter recently. They obtained  constraint on the symmetry energy at twice the saturation density of nuclear matter  as $E_{sym}(2{\rho_{0}})$ = $39.2^{+12.1}_{-8.2}$ MeV at 68\% credible level.

Motivated by the above interesting works, we will investigate the constraint on the nuclear symmetry energy and some of the properties of canonical neutron star through Bayesian inference based on the distribution of tidal deformabilities and component masses of GW170817 in this work.

The paper is organized as follows. In the next section, the isospin-dependent parametric EOS for dense neutron-rich nucleonic matter and properties of neutron star are outlined.
In Sec. 3, through performing Bayesian analysis by correlating the EOS with the GW170817 data released by LIGO and VIRGO, the posterior distribution of the parameter space of the EOS and the symmetry energy of the super dense nuclear matter are presented. Then, In Sec. 4, we present the constraint on the radii and tidal deformabilities of a canonical neutron star through the corresponding posterior distribution. A brief summary is given at the end.

\section{Isospin-dependent parametric EOS and neutron star properties}

Here we give  a brief outline of the isospin-dependent parametric EOS, where the dense nuclear matter is supposed to be composed of neutrons, protons, electrons and muons at $\beta$-equilibrium and charge neutral \cite{ZNB 1,Xie2019}.

The energy density of dense nuclear matter with isospin asymmetry $\delta=(\rho_n-\rho_p)/\rho$ at density $\rho$ can be expressed as
\begin{equation} \label{EQ0}
\epsilon(\rho,\delta) = \rho[E(\rho,\delta)+M_{N}] + \epsilon_{l}(\rho,\delta),
\end{equation}
where $M_{N}$ is the average nucleon mass ($M_{N}$ = 939 MeV), $\epsilon_{l}(\rho,\delta)$ is the lepton energy density, and $E(\rho,\delta)$ is the nucleon specific energy. The pressure of dense nuclear matter can be calculated by
\begin{equation} \label{EQ1}
P(\rho,\delta)={\rho}^2\frac{d\epsilon(\rho,\delta)/{\rho}}{d\rho}.
\end{equation}
The nucleon specific energy $E(\rho,\delta)$ for neutron-rich nuclear matter can be well approximated by the empirical parabolic law as \cite{Bombaci, Li}
\begin{equation} \label{EQ2}
E(\rho,\delta)=E_0(\rho)+E_{\textrm{sym}}(\rho)\cdot\delta^2+O(\delta^4),
\end{equation}
where ${E_{0}(\rho)}$ and $E_{\textrm{sym}}(\rho)$ are the energy in symmetric nuclear matter and the symmetry energy of asymmetric nuclear matter, respectively. They can be parameterized by the following equations \cite{ZNB 1}
\begin{equation} \label{EQ3}
E_0(\rho)=E_0(\rho_0)+\frac{K_0}{2}(\frac{\rho-\rho_0}{3\rho_0})^2+\frac{J_0}{6}(\frac{\rho-\rho_0}{3\rho_0})^3,
\end{equation}

\begin{equation} \label{EQ4}
E_{\textrm{sym}}(\rho)=E_{\textrm{sym}}(\rho_0)+L(\frac{\rho-\rho_0}{3\rho_0})+\frac{K_{\textrm{sym}}}{2}(\frac{\rho-\rho_0}{3\rho_0})^2+\frac{J_{\textrm{sym}}}{6}(\frac{\rho-\rho_0}{3\rho_0})^3,
\end{equation}
where $\rho_{0}$ is the nuclear saturation density. According to the researches near nuclear saturation density, the most probable values of parameters in Eqs. (\ref{EQ3}) and Eqs. (\ref{EQ4}) are as follows: $K_{0}$ = 240 $\pm$ 20 MeV, $E_{\textrm{sym}}(\rho_{0})$ = 31.7 $\pm$ 3.2 MeV, $L$ = 58.7 $\pm$ 28.1 MeV, and $-$300 $\leq$ $J_{0}$ $\leq$ 400 MeV, $-$400 $\leq$ $K_{\textrm{sym}}$ $\leq$ 100 MeV, $-$200 $\leq$ $J_{\textrm{sym}}$ $\leq$ 800 MeV \cite{Shlomo, Piekarewicz, Li 2, ZNB 2, Li 3, Oertel}. It is shown that the parameters $K_{0}$, $E_{\textrm{sym}}(\rho_{0})$ and $L$ have already been constrained in a very narrow range, while $J_{0}$, $K_{\textrm{sym}}$ and $J_{\textrm{sym}}$ have large uncertainties. To simplify the calculation, here we choose the most probable values for $K_{0}$, $E_{\textrm{sym}}(\rho_{0})$ and $L$ as $K_{0}$ = 240 MeV, $E_{\textrm{sym}}(\rho_{0})$ = 31.7 MeV and $L$ = 58.7 MeV. For more details about this EOS please refer to Ref. \cite{ZNB 1}.  Through varying the parameters $J_{0}$, $K_{\textrm{sym}}$ and $J_{\textrm{sym}}$ within their allowed ranges, we can generate sufficiently large number of EOS to perform the Bayesian analysis. 
Compared  with the multisegment polytropic EOS, the parametric EOS model builds a more convenient way to extract the symmetry energy of the asymmetric nuclear matter from the astronomical observations.   In this work, the core matter of neutron star is described by the parametric EOS model, while the inner crust and the outer crust of neutron star are described by the NV EOS model \cite{Negele} and BPS EOS model \cite{Baym}, respectively. We choose resolution for EOS tables as Ref. \cite{chen}.

The structure of neutron star is governed by Tolman-Oppenheimer-Volkoff (TOV) equations \cite{Tolman,Oppenheimer}

\begin{equation} \label{EQ5}
\frac{dP}{dr}=-\frac{G[m(r)+4{\pi}{r}^3{P(r)/c^{2}}]{[\epsilon(r)+P(r)]}}{{c^{2}{r}{[r-2G{m(r)/c^{2}}]}}},
\end{equation}

\begin{equation} \label{EQ6}
\frac{dm}{dr}={4\pi{r}^2\epsilon(r)}/{c^{2}},
\end{equation}
where $\epsilon(r)$ and $P(r)$ are the energy density and pressure at radius $r$, $m(r)$ denotes the mass enclosed within radius $r$, $G$ is the gravitational constant and $c$ is the speed of light. For a given EOS, the TOV equations can be numerically integrated from the origin $(r=0)$ to the surface $(r=R)$, where the pressure vanishs, to obtain the $M$-$R$ relation of neutron star.

The  tidal deformability describes how a neutron star deforms under an external gravitational field produced by its companion star. It can be given by \cite{Carson,Flanagan,Damour2009,Damour2010,Hinderer2008,Hinderer2010,Yagi2013}

\begin{equation} \label{EQ7}
\Lambda=\frac{2}{3}k_{2}(\frac{c^{2}}{G}\frac{R}{M})^{5},
\end{equation}
where the $k_{2}$ denotes the second tidal Love number which has to be solved together with the TOV equations  \cite{Damour2009}.

\begin{figure}[!htb]
	
	\centering
		\includegraphics[height=10.0cm,width=16cm]{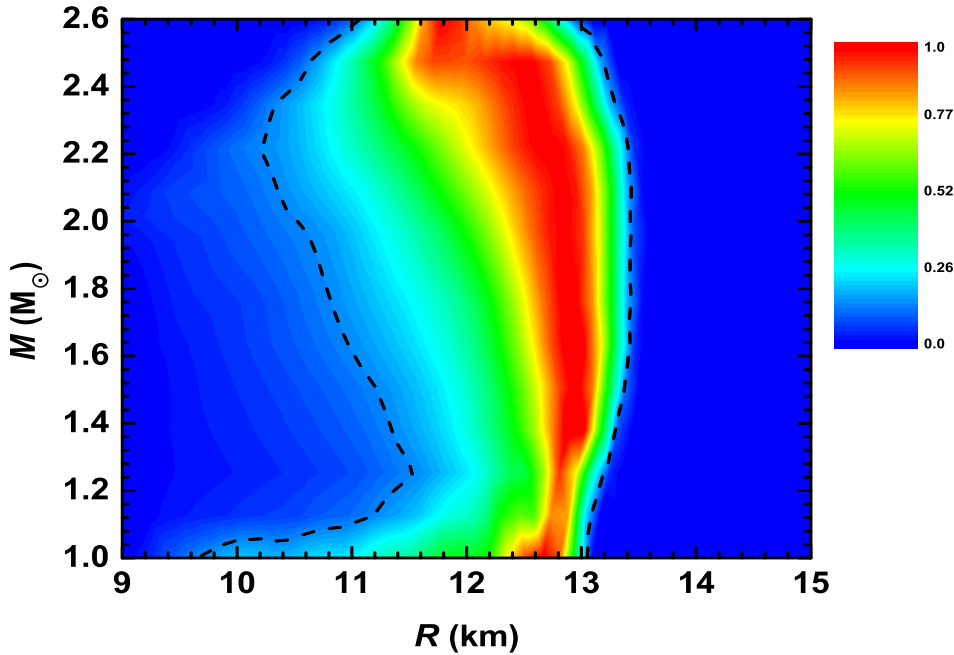}
		\caption{The prior distribution of the $M$-$R$ relation, where the color from red to blue indicates the probability density from high to low. The black dash line denotes the 90\% credible interval.}
	
\label{Fig.1}
\end{figure}

\begin{figure}[!htb]

\centering

\includegraphics[height=10.0cm,width=16cm]{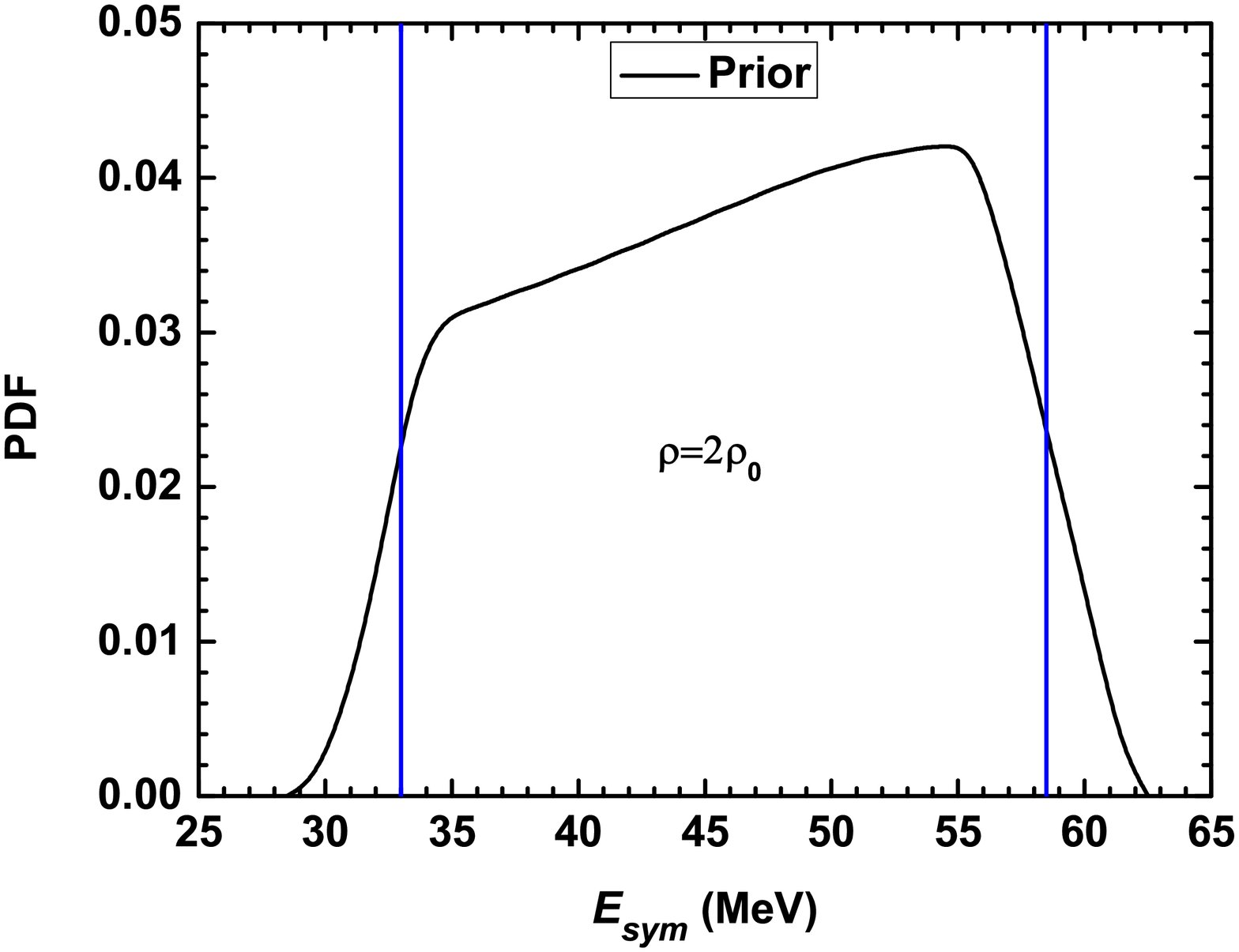}

 \caption{The prior distribution of $E_{\textrm{sym}}{(2\rho_{0})}$, where the solid vertical lines represent the 90\% credible interval for $E_{\textrm{sym}}{(2\rho_{0})}$. The PDF is short for probability density function.}	
\label{Fig.2}

\end{figure}

\begin{figure}[!htb]
	\centering
	\subfigure[]
	{
		\begin{minipage}{16cm}
			\centering
			\includegraphics[scale=0.46]{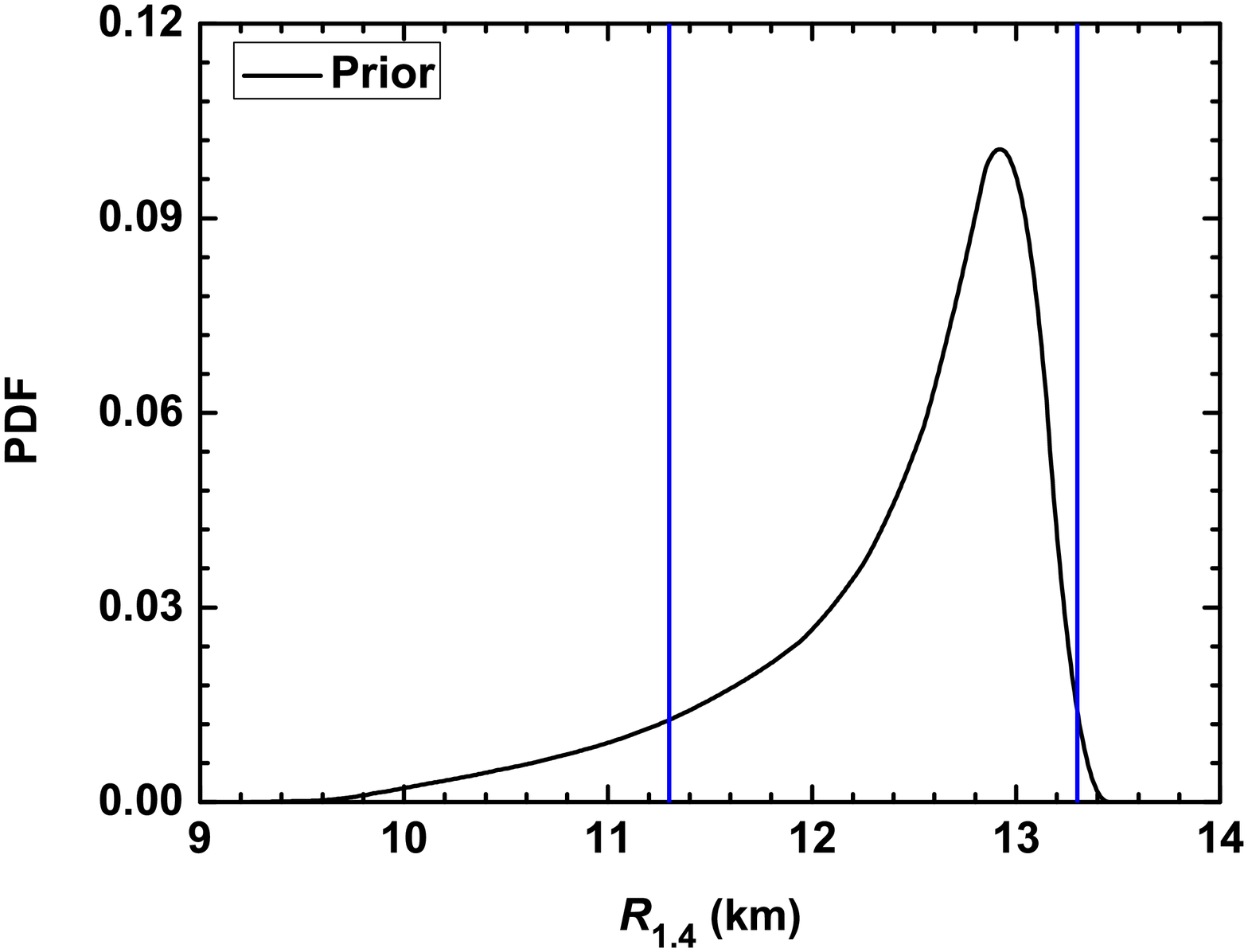}
		\end{minipage}
	}
	\subfigure[]
	{
		\begin{minipage}{16cm}
			\centering
			\includegraphics[scale=0.46]{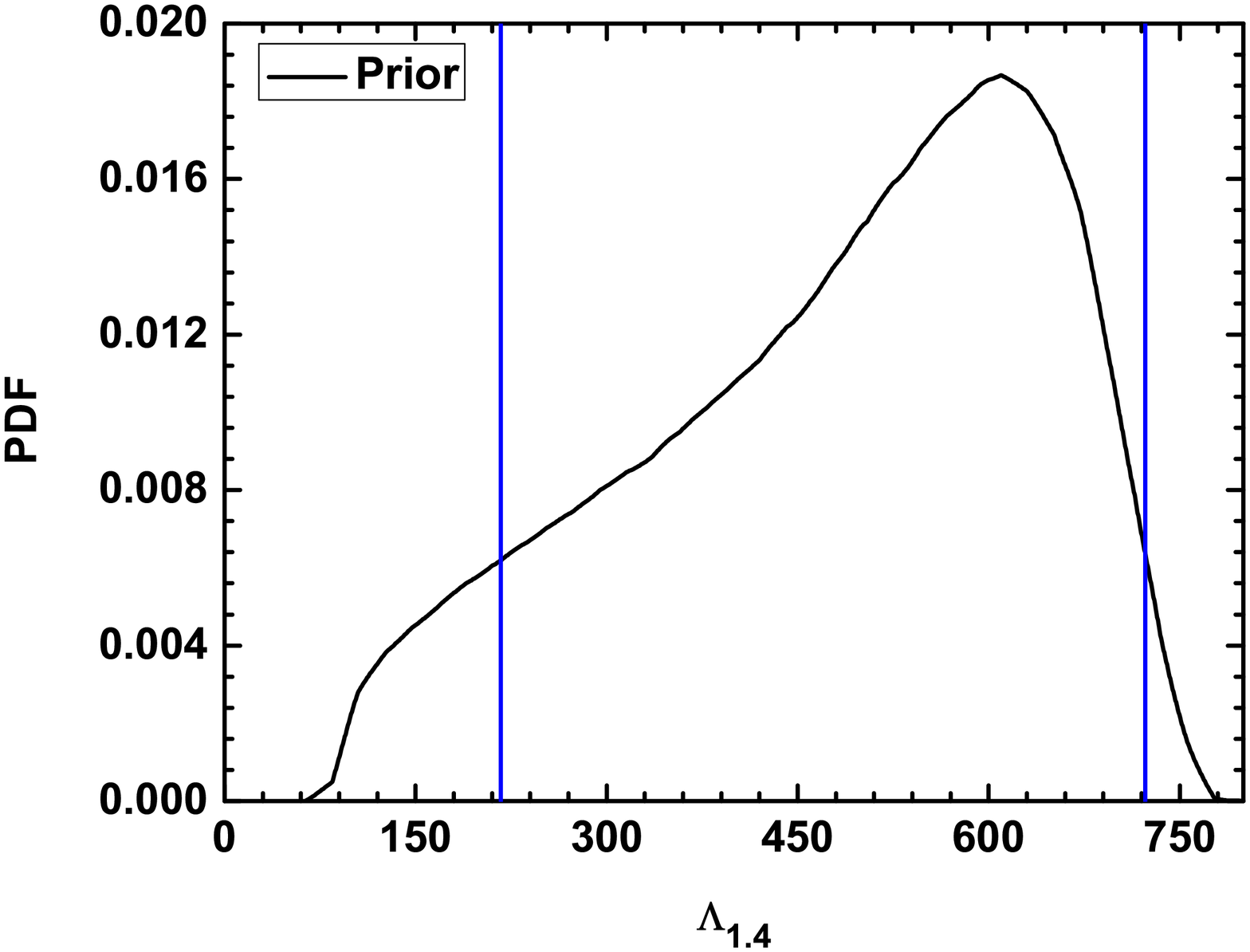}
		\end{minipage}
	}
	\caption{The prior distribution of $R_{1.4}$ (a) and $\Lambda_{1.4}$ (b), where $R_{1.4}$ and $\Lambda_{1.4}$ denote the radii and the dimensionless tidal deformabilities of  canonical neutron stars (with mass of 1.4 $M_{\odot}$), and the solid vertical lines represent the 90\% credible interval for $R_{1.4}$ and $\Lambda_{1.4}$, respectively.}
\label{Fig.3}
\end{figure}

From terrestrial experiments, there are rudimentary constraints on $K_{\textrm{sym}}$, $J_{\textrm{sym}}$, and $J_{0}$  \cite{Tews2017}.
One of our prior assumptions is independent and uniform in their parameter space. In this work, we use Monte Carlo random sampling method to generate two million EOS. Then we screen the generated EOS by causality and by supporting the recently observed heaviest stellar mass 2.14 $M_{\odot}$ of neutron star J0740+6620 \cite{Cromartie}. The remaining EOS  are about 1.6 million.

Calculating the TOV equations by inputting the remaining  EOS, we obtain the prior distribution of the $M$-$R$ relation, as shown in Fig. \ref{Fig.1}.
It is shown that most of the radii of neutron stars with lower mass ($1.0 - 1.2 ~M_{\odot}$) are  concentrated  in a relatively narrow range of $12.4-12.8$ km (the red region), as presented in Fig. \ref{Fig.1}.  This prior distribution means that the difference of the parameters $K_{\textrm{sym}}$, $J_{\textrm{sym}}$ and $J_{0}$ has a relatively weak effect on the radius of neutron star with lower mass.
It is worth noting that the prior distribution reflects the general features of the parametric EOS, but not including the impact from GW170817.
According to  Eqs. (\ref{EQ4}), we can obtain the  prior probability density of $E_{\textrm{sym}}{(2\rho_{0})}$, as shown in Fig. \ref{Fig.2}. Within 90\% credible level, the prior probability of $E_{\textrm{sym}}(2{\rho_{0}})$ is constrained in a range of $E_{\textrm{sym}}(2{\rho_{0}})$ = $54.5^{+4.0}_{-21.5}$ MeV. 

Normally, we call a neutron star with mass of 1.4 $M_{\odot}$ as canonical neutron star as most observed neutron stars have stellar masses near 1.4 $M_{\odot}$ \cite{website 1}. There have been massive researches on the properties of canonical neutron star in recent years \cite{Wen2012,wen,Lattimer2014,Jiang2019,Lattimer}, especially after the GW170817 event \cite{Abbott a,Abbott d,Abbott 6,Lim,Zhao2018,Raithel2018,Fattoyev2018,Annala2018,Bauswein2017,Tews2018}. Here we will also focus on the properties of canonical neutron stars. In Fig. \ref{Fig.3}, we show the prior distribution of $R_{1.4}$ (a) and $\Lambda_{1.4}$ (b),  where $R_{1.4}$ and $\Lambda_{1.4}$ denote the radius and the dimensionless tidal deformability of a canonical neutron star, respectively. It is shown that in the prior distribution and within 90\% credible level, the radius  is distributed  in a range of 11.30 km $\leq$ $R_{1.4}$ $\leq$ 13.30 km and the dimensionless tidal deformability is distributed in a range of 217 $\leq$ $\Lambda_{1.4}$ $\leq$ 723,
where the most probable value of $R_{1.4}$ is 12.9 km and  the most probable value of $\Lambda_{1.4}$ is 620, respectively.
The relatively large value of the expected $E_{\textrm{sym}}{(2\rho_{0})}$, radius and tidal deformability indicate that the prior results, which are consistent with the former studies \cite{Lattimer2014,Nature2006,Psaltis2009,Cabrera2010,Psaltis2016,Suleimanov2011,Steiner2006,Steiner2013,Guillot2013,Guillot2014,Bogdanov2016},  prefer a stiffer EOS for a neutron star.

\section{Constraint on the parameter space of EOS and the symmetry energy at twice saturation density}

\begin{figure}[!htb]
	\centering
	\subfigure[]
	{
		\begin{minipage}{7cm}
			\centering
			\includegraphics[scale=0.27]{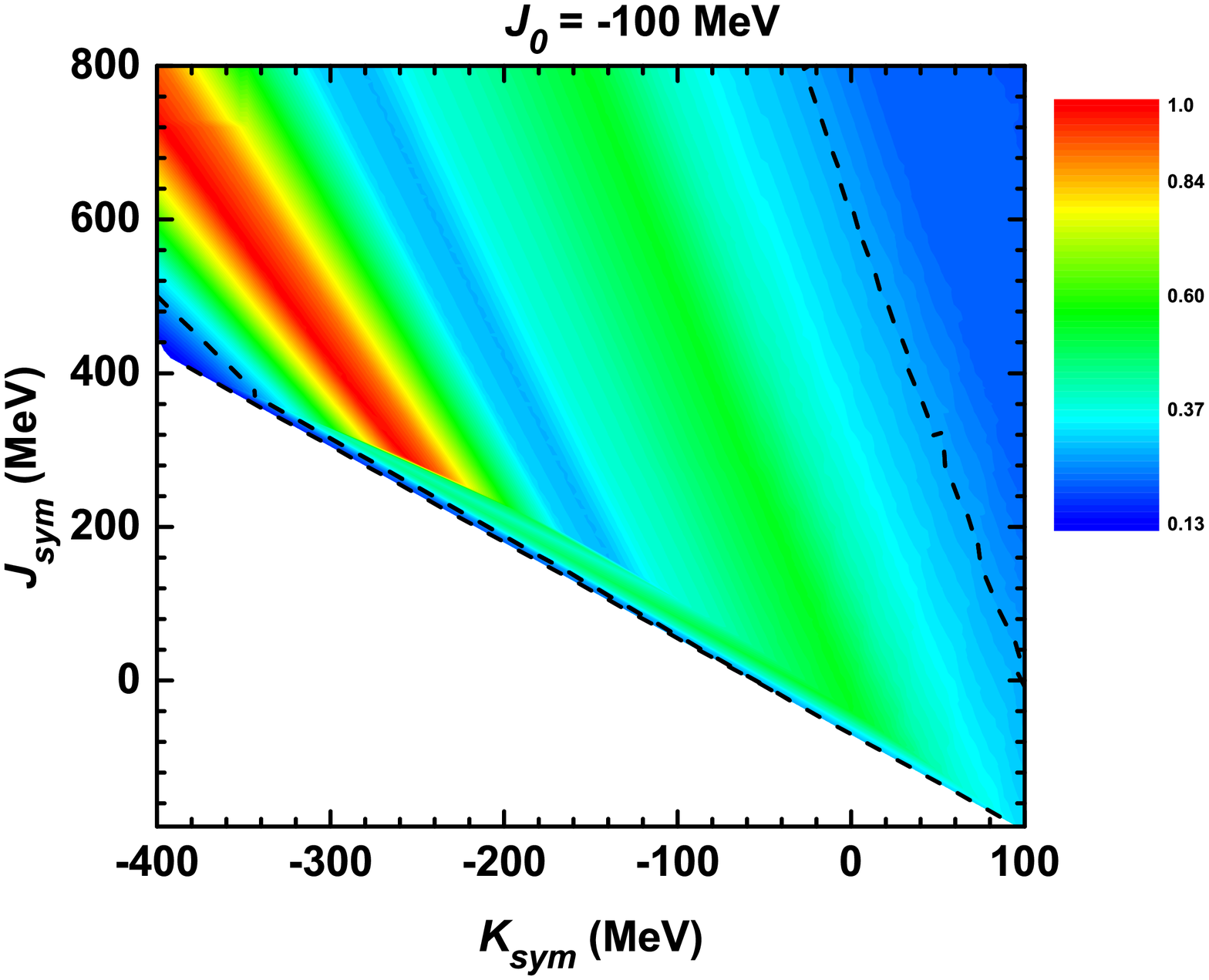}
	\end{minipage}
    }
	\subfigure[]
	{
		\begin{minipage}{7cm}
			\centering
			\includegraphics[scale=0.27]{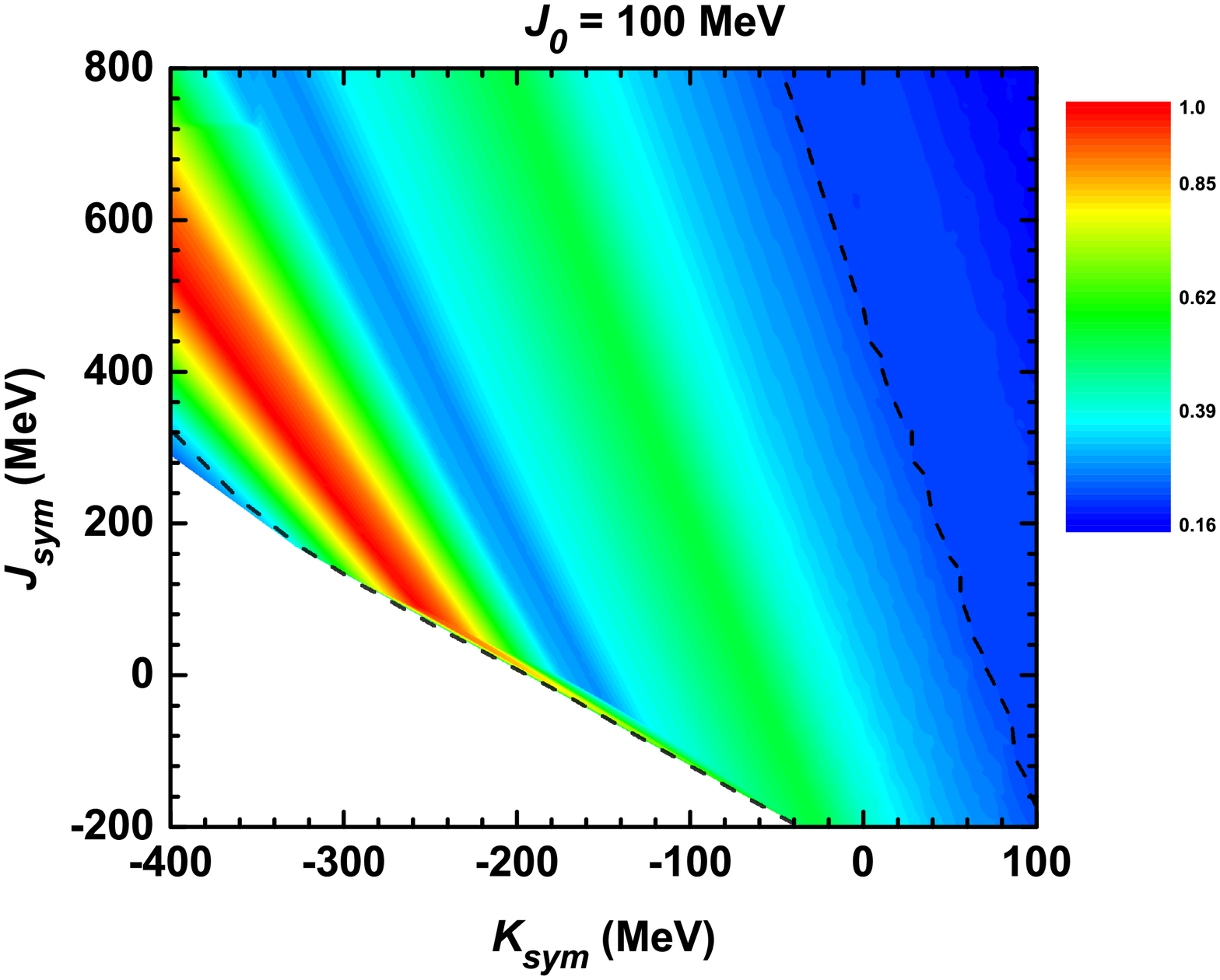}
		\end{minipage}
	}
	\subfigure[]
{
	\begin{minipage}{7cm}
		\centering
		\includegraphics[scale=0.27]{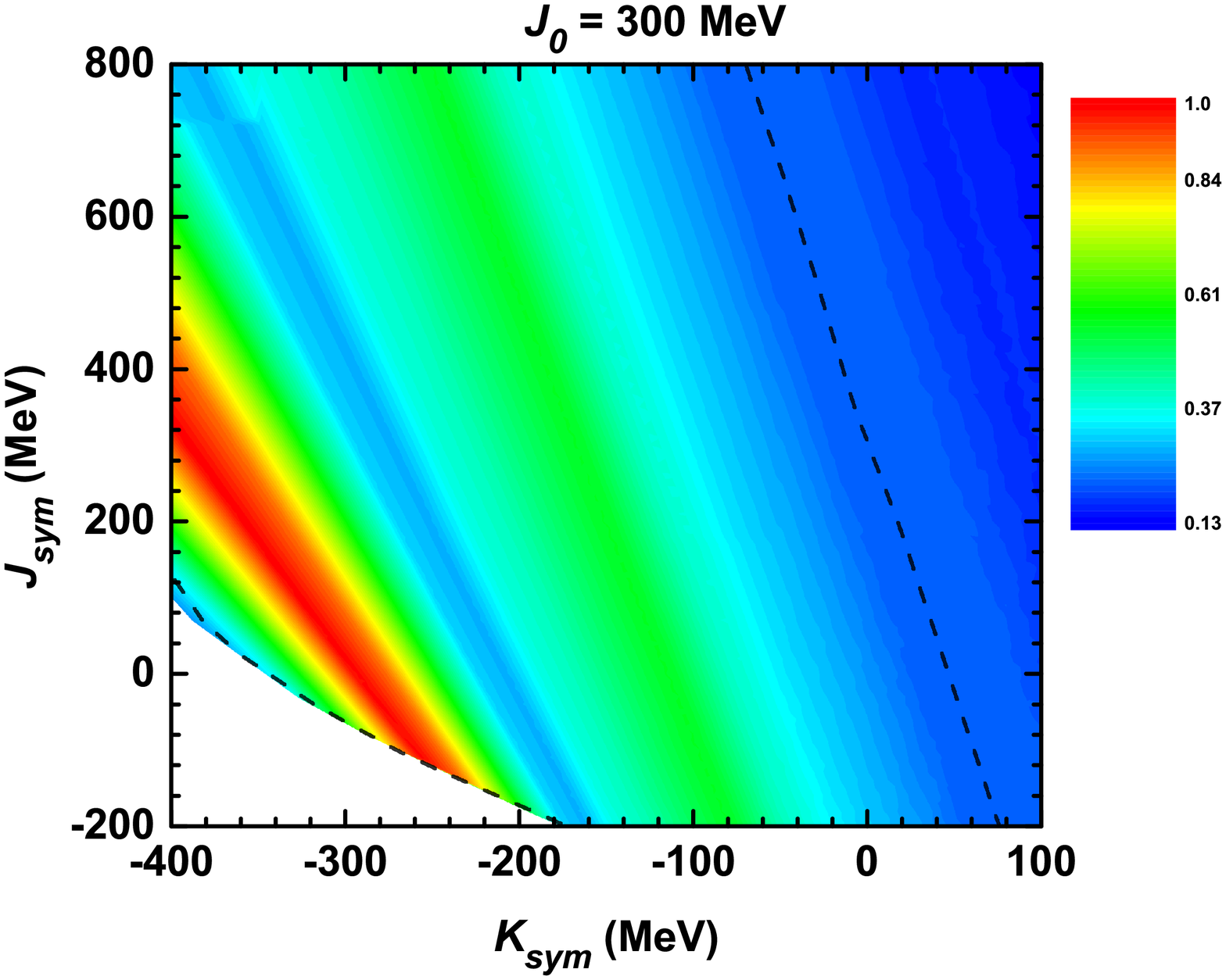}
	\end{minipage}
}
	\subfigure[]
{
	\begin{minipage}{7cm}
		\centering
		\includegraphics[scale=0.27]{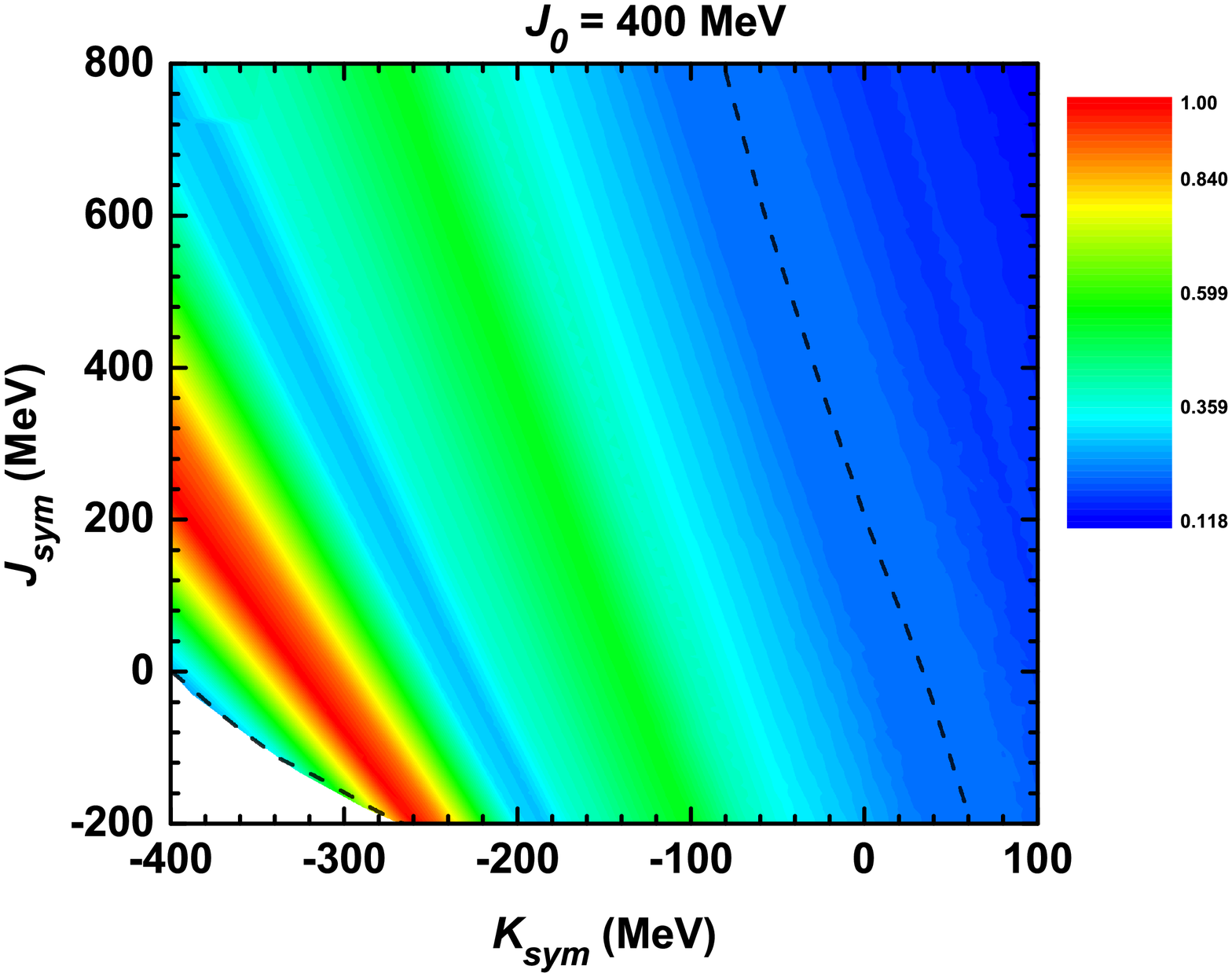}
	\end{minipage}
}
	\caption{The posterior probability distribution in parameter space, where the color from red to blue indicates the probability density from high to low. The white areas are the forbidden parameter spaces, where the maximum mass $M_{max}$=2.14 $M_{\odot}$ can not be supported. The black dash line denotes the 90\% credible interval.}
\label{Fig.4}	
\end{figure}

\begin{figure}[!htb]
	\centering
	\subfigure[]
	{
		\begin{minipage}{7cm}
			\centering
			\includegraphics[scale=0.27]{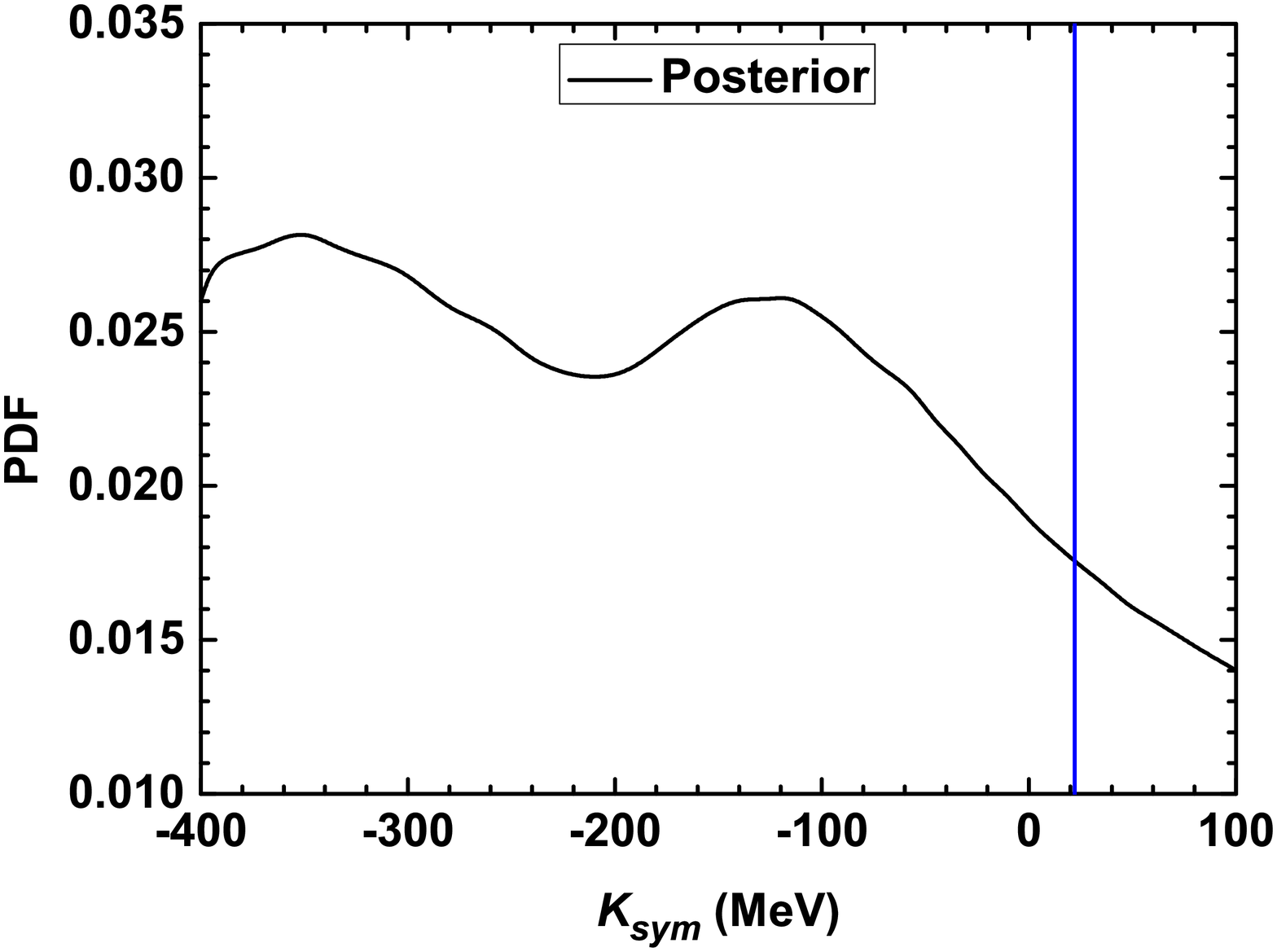}
		\end{minipage}
	}
	\subfigure[]
	{
		\begin{minipage}{7cm}
			\centering
			\includegraphics[scale=0.27]{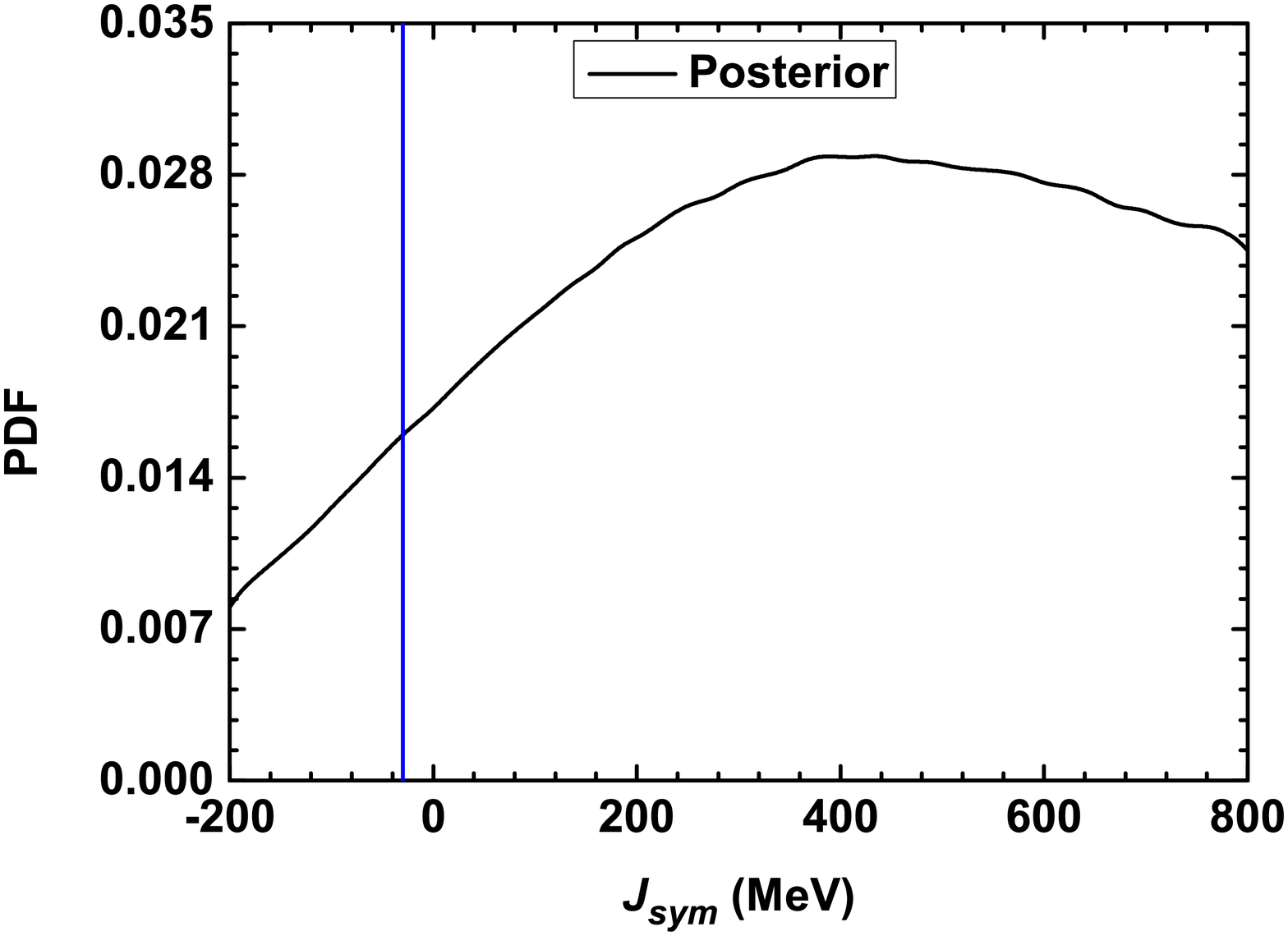}
		\end{minipage}
	}
	\subfigure[]
	{
		\begin{minipage}{7cm}
			\centering
			\includegraphics[scale=0.27]{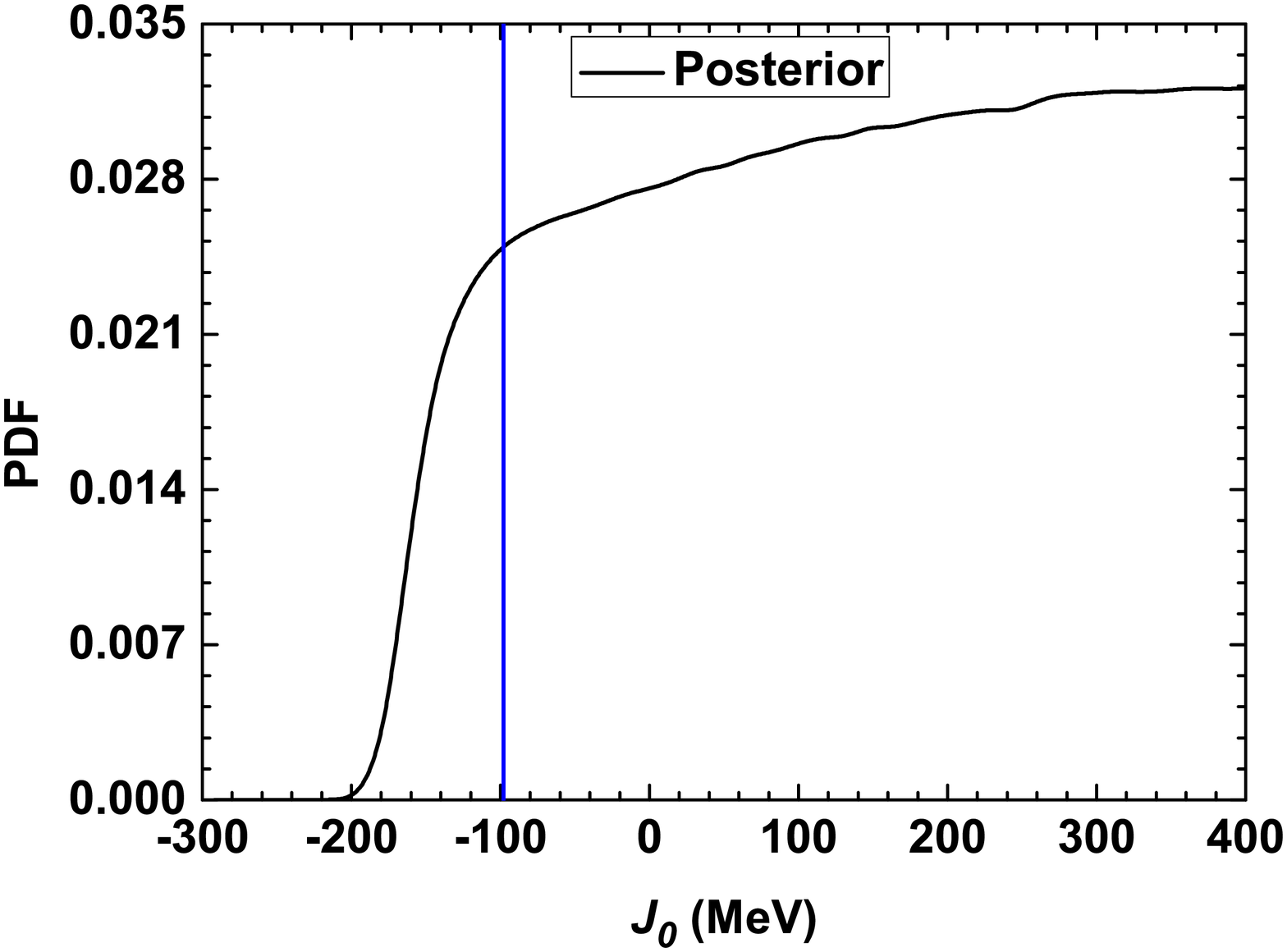}
		\end{minipage}
	}
	\caption{The posterior distribution of $K_{\textrm{sym}}$ , $J_{\textrm{sym}}$ and $J_{0}$, where the solid vertical lines represent the 90\% credible interval for $K_{\textrm{sym}}$ , $J_{\textrm{sym}}$ and $J_{0}$, respectively.}
	\label{Fig.5}	
\end{figure}

\begin{figure}[!htb]
	
	\centering
	
	\includegraphics[height=10.0cm,width=16cm]{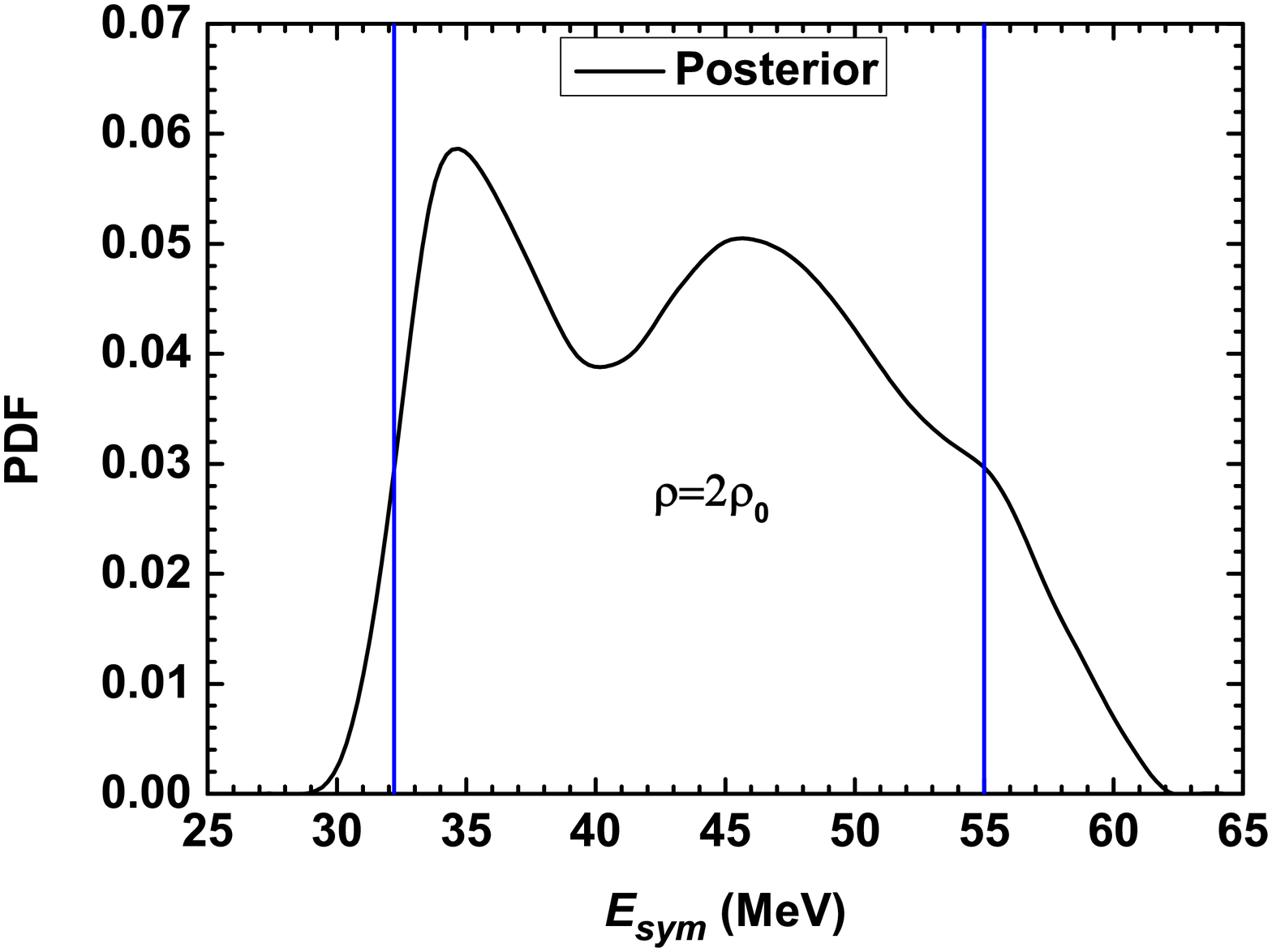}
	
	\caption{The posterior distribution of $E_{\textrm{sym}}{(2\rho_{0})}$, where the solid vertical lines represent the 90\% credible interval for $E_{\textrm{sym}}{(2\rho_{0})}$.}
\label{Fig.6}	
\end{figure}

Based on the distribution of stellar masses and tidal deformabilities of the binary neutron star in GW170817 \cite{Abbott 6},  we define the  likelihood function of each set of parameters of EOS and perform Bayesian inference to analyze the impact on the parameter space of EOS and the symmetry energy. For the details of the Bayesian inference approach adopted in this work please see the Appendix.

For the convenience to show the posterior probability of the EOS parameters, we plot the posterior probability of $K_{\textrm{sym}}$ and $J_{\textrm{sym}}$ in a two-dimensional diagram by fixing the parameter $J_{0}$, as shown in Fig. \ref{Fig.4}, where four values of $J_{0}$: (a) $-$100 MeV, (b) 100 MeV, (c) 300 Mev and (d) 400 MeV are adopted. From Fig.  \ref{Fig.4}, we can see that the higher probability density area is  located in the left area. Moreover, it is also shown that for a given $J_{0}$, lower values of $J_{\textrm{sym}}$ and $K_{\textrm{sym}}$ are preferred to support the observed data of GW170817. From the definitions of the parameters $K_{\textrm{sym}}$, $J_{\textrm{sym}}$ and $J_{0}$ in Eqs. (\ref{EQ3}) and Eqs. (\ref{EQ4}) and according to Eq. (\ref{EQ2}), it is easy to understand that higher values of the set of parameters correspond with a stiffer EOS. Therefore, the posterior probability of the parameters of EOS as shown in Fig. \ref{Fig.4} indicates that the observation of GW170817 prefers a relatively softer EOS. In addition, the white areas at the left bottom in Fig. \ref{Fig.4} are forbidden areas, where the related EOS can not support the maximum mass of $M_{max}$=2.14 $M_{\odot}$. The black dash line denotes the boundary of 90\% credible interval.

 In addition, according to posterior probability distribution of parameter space , we can get a roughly linear constraint of the parameter space (from the boundary of 90\% credible interval). That is, the upper boundary (located in the right area) can be approximately expressed as
\begin{equation}\label{EQ8}
 8.06K_{\textrm{sym}} + 1.21J_{\textrm{sym}} + J_{0}  \leq 676.25~ \textrm{MeV},
\end{equation}
and the lower boundary (located in the left area) can be approximately expressed as
\begin{equation}\label{EQ9}
1.42K_{\textrm{sym}} + 1.04J_{\textrm{sym}} + J_{0}  \geq -177.00~ \textrm{MeV},
\end{equation}
where $K_{\textrm{sym}}$, $J_{\textrm{sym}}$, and $J_{0}$ are the values of corresponding parameters in unit of MeV. The two constraints can reduce the parameter space to about 50\% of the original parameter space. 

The posterior distributions of $K_{\textrm{sym}}$ , $J_{\textrm{sym}}$ and $J_{0}$ are presented in Fig. \ref{Fig.5}. Within 90\% credible level, it is shown that the parameters $K_{\textrm{sym}}$ , $J_{\textrm{sym}}$ and $J_{0}$  are constrained in $-400.0$ MeV $< K_{\textrm{sym}} < 22.0$ MeV, $-30.0$ MeV $< J_{\textrm{sym}} < 800.0$ MeV and $-98.0 $ MeV $< J_{0} < 400.0$ MeV, respectively.

It is well known that the radius of normal neutron star is essentially determined by the pressure  around the twice saturation density ($2\rho_{0}$) of nuclear matter \cite{Lattimer}. Thus the knowledge of the EOS, especially the symmetry energy around the twice saturation density is very important in
understanding the radius of neutron star. Up to now, a lot of research works have been done to constrain the symmetry energy at twice the saturation density \cite{Xie2019,wen,ZNB 3,ying,Roca-Maza,Dutra2012,Dutra2014,Vidana,Z. H. Li,Sammarruca,Akmal,Friedman,Wiringa1988,Sammarruca2014,He,T}.
For example, through employing three sets of observational related radii data and three sets of imaginary radii data of canonical neutron star to perform Bayesian analysis, Xie and Li \cite{Xie2019} inferred the nuclear symmetry energy $E_{\textrm{sym}}{(\rho)}$ by the parametric EOS and constrained the symmetry energy at twice the saturation density as $E_{\textrm{sym}}(2{\rho_{0}})$ = $39.2^{+12.1}_{-8.2}$ MeV at 68\% credible level.
 Based on the oscillation modes of canonical neutron stars, Wen $et ~ al.$ \cite{wen}  predicted that the symmetry energy at twice the saturation density $E_{\textrm{sym}}(2{\rho_{0}})$  should be in a range of $54.5^{+6.5}_{-6.5}$ MeV if the frequency of f-mode takes a value of $f$ = $1.640^{+0.016}_{-0.016}$ kHz, while  $E_{\textrm{sym}}(2{\rho_{0}})$ should be in a range of $43.0^{+6.5}_{-6.5}$ MeV if the frequency of f-mode takes a value of $f$ = $1.800^{+0.018}_{-0.018}$ kHz.
 By comprehensively combining the observational constraints on the radius, maximum mass, tidal deformability and causality condition of neutron stars, Zhang and Li \cite{ZNB 3} deduced that the symmetry energy at twice the saturation density should be in a range of $E_{\textrm{sym}}(2{\rho_{0}})$ = 46.9 $\pm $10.1 MeV.
  Very recently, by using the eSHF model to simultaneously analyze the data from terrestrial nuclear experiments and astrophysical
observations based on strong interaction, electromagnetic and gravitational measurements, Zhou $et ~al.$ \cite{ying}  derived the symmetry energy at twice the saturation density  $E_{\textrm{sym}}(2{\rho_{0}})$  in a range of  [$39.4^{+7.5}_{-6.4}$, $54.5^{+3.1}_{-3.2}$] MeV, respectively. To sum up,  current studies constrain the symmetry energy at twice the saturation density  in a range of [30, 60] MeV.

 By employing the posterior distribution of the parameter space and Eqs. (\ref{EQ4}), we obtain the  posterior probability density of $E_{\textrm{sym}}{(2\rho_{0})}$, as shown in Fig. \ref{Fig.6}. It is shown that the symmetry energy at twice the saturation density of nuclear matter is constrained in a range of $E_{sym}(2{\rho_{0}})$ = $34.5^{+20.5}_{-2.3}$ MeV at 90\% credible level. Obviously, there is a big difference of the maximum probability point between the prior and posterior distribution. The latter one prefers a relatively low symmetry energy. Moreover, both of the prior and posterior distribution of $E_{sym}(2{\rho_{0}})$ are consistent with the conclusions of above literatures.

\section{The constraint on the radii and tidal deformabilities through the GW170817}

\begin{figure}[!htb]
	
	\centering
	
	\includegraphics[height=10.0cm,width=16cm]{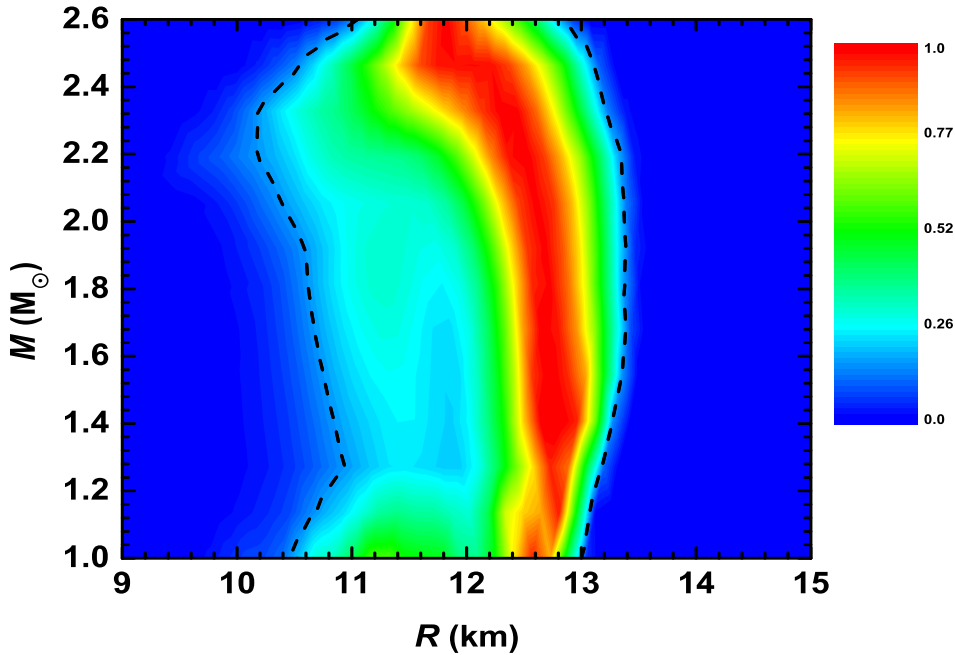}
	
	\caption{The posterior distribution of $M$-$R$ relation, where the color from red to blue indicates the probability density from high to low. The black dash line denotes 90\% credible interval.}
\label{Fig.7}	
\end{figure}

\begin{figure}[!htb]
	\centering
	\subfigure[]
	{
		\begin{minipage}{16cm}
			\centering
			\includegraphics[scale=0.46]{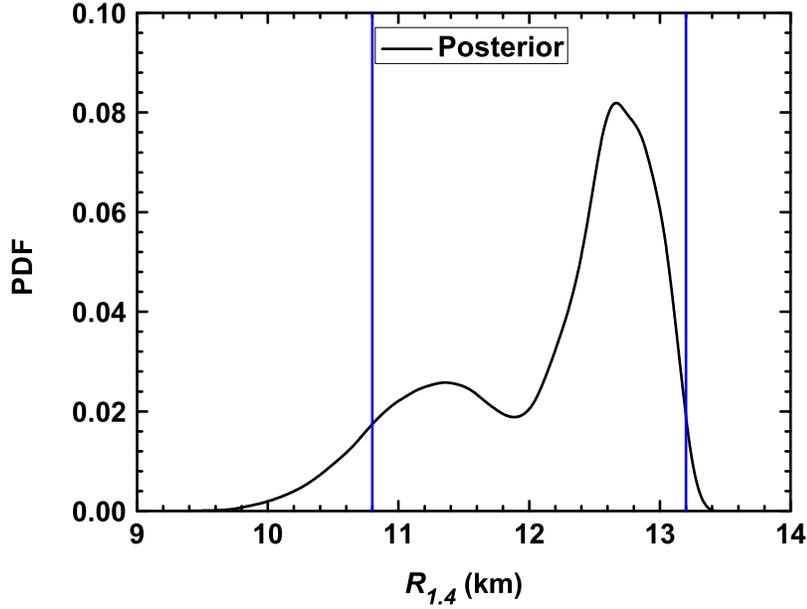}
		\end{minipage}
	}
	\subfigure[]
	{
		\begin{minipage}{16cm}
			\centering
			\includegraphics[scale=0.46]{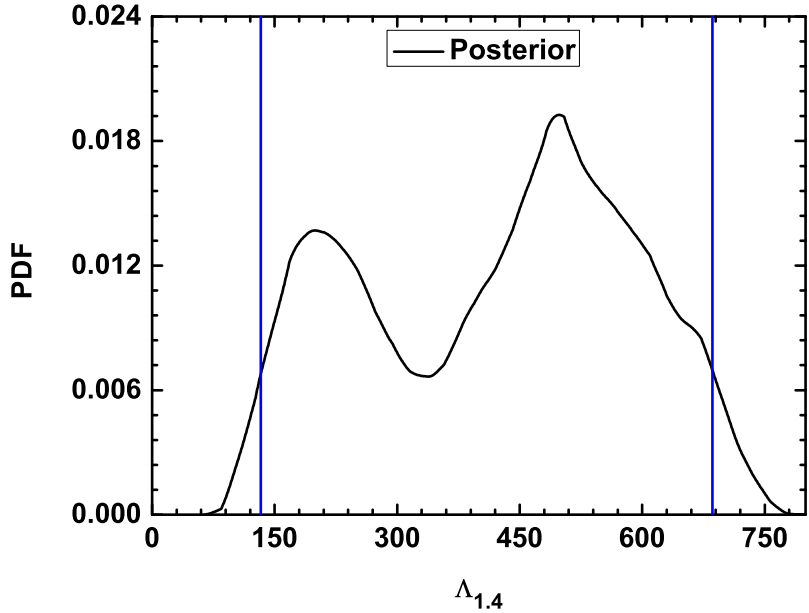}
		\end{minipage}
	}
	\caption{The posterior distribution of radii $R_{1.4}$ (a) and dimensionless tidal deformabilities $\Lambda_{1.4}$ (b) of  canonical neutron stars, where the solid vertical lines represent the 90\% credible interval for $R_{1.4}$ and $\Lambda_{1.4}$, respectively.}
	\label{Fig.8}
\end{figure}

For comparison with prior distribution of the $M$-$R$ relation, here we present the corresponding posterior distribution in Fig. \ref{Fig.7}. We use Monte Carlo random sampling method (by the probability density of posterior parameter space) to generate two million EOS, and then $M$-$R$ relation, $R_{1.4}$, and $\Lambda_{1.4}$ are calculated statistically. Of course, according to the prior assumptions, it can be known that the EOS generated by posterior parameter space have ensured constraints mentioned in section III. The distribution of posterior $M$-$R$ relation is shown in Fig. \ref{Fig.7}. It can be seen from Fig. \ref{Fig.7} that, comparing with the prior $M$-$R$ relation distribution, the posterior $M$-$R$ relation distribution spreads to the left, especially at high stellar mass. So the posterior $M$-$R$ relation prefers a softer EOS.

The posterior distribution of the  radii $R_{1.4}$ and the dimensionless tidal deformabilities $\Lambda_{1.4}$ of canonical neutron stars are presented  in Fig. \ref{Fig.8} (a) and (b), respectively.  Unlike the prior distribution (see Fig. \ref{Fig.3}), there are two peaks in the posterior distribution of $R_{1.4}$ and $\Lambda_{1.4}$ in Fig. \ref{Fig.8}. We notice that in Fig. \ref{Fig.4}, each subgraph has two higher probability density areas of the  posterior probability in the EOS parameter space and one of them has a relatively lower probability density. Considering a certain relevance between the tidal deformability and the radius, it is easy to understand that  the  two peaks in Fig. \ref{Fig.8}  are consistent with the two higher probability density areas in  Fig. \ref{Fig.4}.

According to our calculations, at 90\% credible level in the posterior distribution, the radius of a canonical neutron star is distributed  in a range of $R_{1.4}$ = $12.6_{-1.8}^{+0.6}$ km and its dimensionless tidal deformability is distributed in a range of  $\Lambda_{1.4}$ = $500_{-367}^{+186}$.  Comparing with the  prior distribution of radius ($R_{1.4}$ = $12.9_{-1.6}^{+0.4}$ km) and the tidal deformability ($\Lambda_{1.4}$ = $620_{-403}^{+103}$ ), we can find that the most probable value of the radius and the tidal deformability of the posterior distribution is  smaller than that of the prior distribution, which means that the posterior distribution  prefers a relatively softer EOS.

\section{SUMMARY}

The detection of the gravitational waves of the binary neutron star merger event GW170817 provides us important information, such as the distribution of the tidal deformabilities and the  stellar masses of the binary neutron star, to further investigate the properties and the state of matter of neutron stars.
In this work, we investigate the radius and tidal deformability of canonical neutron star and the symmetry energy of the super dense matter through the Bayesian analysis based on the distribution of component masses and tidal deformabilities of binary neutron star merger GW170817 released by LIGO and VIRGO.
To perform the Bayesian analysis, one need to generate a huge number of EOS. Normally, the polytropic EOS model is adopted  to generate the EOS. Normally, the polytropic EOS model is adopted to generate the EOS. Here we adopt the isospin-dependent parametric EOS model  as this kind of model that can provide  a more convenient way to extract the symmetry energy of the asymmetric nuclear matter from the astronomical observations. In this work, two million isospin-dependent parametric EOS are generated by the Monte Carlo random sampling method,  and the generated EOS are further screened by the recently observed heaviest stellar mass 2.14 $M_{\odot}$ of  J0740+6620 and the causality to perform the Bayesian analysis. From this analysis, we find that the parameter space of EOS can be reduced to about 50\% of the original parameter space at 90\% credible level. In the posterior distribution, the symmetry energy at twice the saturation density of nuclear matter can be constrained within $E_{sym}(2{\rho_{0}})$ = $34.5^{+20.5}_{-2.3}$ MeV at 90\% credible level, the radius is distributed  in a range of $R_{1.4}$ = $12.6_{-1.8}^{+0.6}$ km and the dimensionless tidal deformability is distributed in a range of  $\Lambda_{1.4}$ = $500_{-367}^{+186}$ at 90\% credible level.  Comparing with the  prior distribution of $E_{sym}(2{\rho_{0}})$ ($E_{\textrm{sym}}(2{\rho_{0}})$ = $54.5^{+4.0}_{-21.5}$ MeV), radii ($R_{1.4}$ = $12.9_{-1.6}^{+0.4}$ km) and the tidal deformabilities ($\Lambda_{1.4}$ = $620_{-403}^{+103}$ ), one can see that the posterior distribution prefers a softer EOS.

\section{Acknowledgements}
We thank Bao-An Li for helpful discussions. This work is supported by NSFC (Grants No. 11975101 and No. 11722546), Guangdong Natural Science Foundation (Grant No. 2020A151501820) and the talent program of South China University of Technology (Grant No. K5180470). This project has made use of NASA's Astrophysics Data System.

\appendix
\section{Bayesian Inference Approach}

Bayesian statistics give the posterior probability as
\begin{equation} \label{EQ10}
P(\overrightarrow{\theta}|D)=\frac{P(D|\overrightarrow{\theta}){P(\overrightarrow{\theta})}}{\int{P(D|\overrightarrow{\theta})}P(\overrightarrow{\theta})d{\overrightarrow{\theta}}},
\end{equation}
where $P(\overrightarrow{\theta}|D)$ is the posterior probability for the model $\overrightarrow{\theta}$ (isospin-dependent parametric EOS) given the data set D which is the distribution of tidal deformabilities and stellar masses in GW170817, $P(D|\overrightarrow{\theta})$ is the likelihood function for
a given neutron star model $\overrightarrow{\theta}$ to correctly infer the data D; and $P(\overrightarrow{\theta})$ is the prior probability of the model $\overrightarrow{\theta}$ before being correlated with the data D. The denominator in Eq. \ref{EQ10} is the normalization constant, which is a constant for each parametric EOS.

Here the probability distributions of tidal deformabilities and masses (gravitational wave model PhenomPNRT and low spin prior, $\chi$ $\leq$ 0.05) released by LIGO and VIRGO collaboration in GW170817 are employed as the data D. The non-normalized likelihood function is expressed as
\begin{equation} \label{EQ11}
P(D|\overrightarrow{\theta})={\int{dM}\cdot{P_{\Lambda}}\cdot{P_{M}}},
\end{equation}
For the convenience of understanding the construction of the likelihood function, we discuss it in more details as follows. For a given parametric EOS, we can get the properties (mass, tidal deformability) of a series of neutron stars by using our codes. The $P_{\Lambda}$ and $P_{M}$  of each neutron star are derived by combining the tidal deformability and mass of a specified neutron star with the probability distribution of tidal deformabilities and component masses in GW170817. Therefore, for a given parametric EOS, the value of the non-normalized likelihood function can be determined by integrating the Eqs.(\ref{EQ11}).

As we know, one of the tightly constrained physical quantities in GW170817 is the chirp mass, which is defined as
\begin{equation} \label{EQ12}
M_{c}=\frac{(m_{1}{m_{2}})^{3/5}}{(m_{1}+m_{2})^{1/5}},
\end{equation}
where the $m_{1}$ and $m_{2}$ are the component masses of the merged binary neutron stars. The chirp mass in GW170817 is constrained to $M_{c}$ = $1.186^{+0.001}_{-0.001}$ $M_{\odot}$ at 90\% credible level, which is independent of the waveform model  \cite{Abbott 1}.
Here we employ correlated prior for the mass with the specified chirp mass at the most probable value $M_{c}$ = $1.186$ $M_{\odot}$ and completely uncorrelated prior for the deformabilities of binary neutron star in GW170817. Then we use the fixed chirp mass value to set $m_{2}$ for each given $m_{1}$. From this way, we can get the posterior probability of each sample (each parametric EOS).

\end{document}